\documentclass[conference]{IEEEtran}
\pdfoutput=1
\usepackage{amsmath,amssymb,amsfonts}
\usepackage{algorithmic}
\usepackage{graphicx}
\usepackage{textcomp}
\usepackage{xcolor}
\usepackage{hyperref}
\usepackage{enumitem}
\usepackage[style=ieee]{biblatex}
\def\BibTeX{{\rm B\kern-.05em{\sc i\kern-.025em b}\kern-.08em
    T\kern-.1667em\lower.7ex\hbox{E}\kern-.125emX}}
\begin{document}

\title{Federated Single Sign-On and Zero Trust Co-design for AI and HPC Digital Research Infrastructures
}

\author{
    \IEEEauthorblockN{
        Sadaf R. Alam\IEEEauthorrefmark{1}, \and
        Christopher Woods\IEEEauthorrefmark{1}, \and
        Matt Williams\IEEEauthorrefmark{1}, \and
        Dave Moore\IEEEauthorrefmark{1}, \and
        Isaac Prior\IEEEauthorrefmark{1}, \and
        Ethan Williams\IEEEauthorrefmark{1}, \and
        Anna Price\IEEEauthorrefmark{1}, \and
        James Womack\IEEEauthorrefmark{1}, \and
        Simon McIntosh-Smith\IEEEauthorrefmark{1}, \and
        Fan Yang-Turner\IEEEauthorrefmark{1}, \and
        Matt Pryor\IEEEauthorrefmark{2}, \and
        Ilja Livenson\IEEEauthorrefmark{3}
    }
    \and
    \IEEEauthorblockA{
        \IEEEauthorrefmark{1}
        \textit{Bristol Centre for Supercomputing} \\
        \textit{University of Bristol}\\
        Bristol, United Kingdom
    } 
    \and
    \IEEEauthorblockA{
        \IEEEauthorrefmark{2}
        \textit{StackHPC} \\
        Beacon Tower, Colston Street, \\
        Bristol, United Kingdom
    }
    \and
    \IEEEauthorblockA{
        \IEEEauthorrefmark{3}
        \textit{Institute of Computer Science} \\
        \textit{High Performance Computing Center} \\
        \textit{University of Tartu}\\
        Tartu, Estonia
    }
}

\maketitle

\begin{abstract}
Scientific workflows have become highly heterogenous, leveraging distributed facilities such as High Performance Computing (HPC), Artificial Intelligence (AI), Machine Learning (ML), scientific instruments (data-driven pipelines) and edge computing.  As a result, Identity and Access Management (IAM) and Cybersecurity challenges across the diverse hardware and software stacks are growing.  Nevertheless, scientific productivity relies on lowering access barriers via seamless, single sign-on (SSO) and federated login while ensuring access controls and compliance.  We present an implementation of a federated IAM solution, which is coupled with multiple layers of security controls, multi-factor authentication, cloud-native protocols, and time-limited role-based access controls (RBAC) that has been co-designed and deployed for the Isambard-AI and HPC supercomputing Digital Research Infrastructures (DRIs) in the UK.  Isambard DRIs as a national research resource are expected to comply with regulatory frameworks.  Implementation details for monitoring, alerting and controls are outlined in the paper alongside selected user stories for demonstrating IAM workflows for different roles.
\end{abstract}


\section{Introduction}
\label{sec:introduction}
According to UKRI (UK Research Innovation), a Digital Research Infrastructure (DRI) should allow its users to work with data and computation efficiently and securely across all research and innovation challenges\cite{b1}.  Infrastructure, in this context, is defined as: ``Facilities, resources and services that are used by the research and innovation communities to conduct research and foster innovation in their fields. They can include major research equipment (or sets of instruments), knowledge-based resources (such as collections, archives and data), or digital infrastructures (such as data and computing systems and communication networks, as well as any other tools that are essential to achieve excellence in research and innovation).'' These DRIs routinely serve communities such as weather and climate modelling digital twins\cite{b2}, Data and Analytics Research Environments (DARE) UK\cite{b3}, and recently for the safety and security of AI\cite{b4}.  The building blocks of the DRI systems are wide ranging, from large-scale compute facilities to data storage facilities and repositories, as well as stewardship, security, software and skillsets.

Isambard-AI is a new AI Research Resource, or AIRR, for the UK to quickly address the lack of national AI-capable supercomputing facilities for open research, and the impact this was having on AI research, as well as AI-enabled scientific discovery\cite{b5}.  Isambard-AI is based on the HPE Cray EX4000 system, and housed in a new, energy efficient Modular Data Centre in Bristol, UK. Isambard-AI employs 5,448 NVIDIA Grace-Hopper GPUs to deliver over 21 ExaFLOP/s of 8-bit floating point performance for LLM training and inference, and over 250 PetaFLOP/s of 64-bit performance, for under 5MW.  Bristol is deploying another DRI called Isambard-3, which is a national tier-2 HPC platform based on 384 NVIDIA Grace-Grace superchips, delivering over 55,000 CPU cores.  Isambard-AI and Isambard 3 are expected to deliver bare-metal performance for applications that scale to thousands of GPUs and high-speed network endpoints.  Traditional approaches for security relying on multi-tenancy via virtualization and isolation cannot be readily adopted in these environments.  We follow a highly customised approach for multi-tenancy and for supporting the usage of Linux container technologies\cite{b6}.  Many AI and ML users are expected to use, and in many cases are already using, optimized containers such as PyTorch from NGC\cite{b7}.

As a national AI RR alongside the Cambridge Dawn system\cite{b8}, Isambard-AI is expected to provide federated access for research and academia.  At the same time however, it must support the UK’s AI Safety Institute mission, which requires fulfilling regulatory and compliance requirements for access controls and cybersecurity.  These include, but are not limited to, the UK’s national centre for cybersecurity (NCSC) Cyber Assurance Framework (CAF)\cite{b9}. Moreover, we need to align with the UK Government Security Classifications Policy (GSCP) for some use cases for adequate protections against threats\cite{b10}. There are three classification tiers, however, only Official (OFF) is applicable to the Isambard DRIs being a national research resource. These are a set of protective security controls, baseline behaviours and management of tradeoffs and risks for enabling usage in an academic and research environment.

Considering these competing requirements of lowering access barriers, bare-metal performance and a prescriptive cybersecurity posture, we follow the guidelines outlined in the NIST SP 800-223 document ``High-Performance Computing Security: Architecture, Threat Analysis, and Security Posture''\cite{b11}. Specifically, we follow the concept of ``zones'', physically and logically separating the Access, Management and High Performance Computing zones, and in some cases adding granularity and layers for roles and access controls.

The Isambard team primarily adopts two design patterns from the outset: federation and zero trust.  The federation model that has been used relies on the level of assurance and trust (LoA and LoT) that have been introduced by AARC2\cite{b12}.  One of its instantiations is MyAccessID\cite{b13}. Similar services include Globus auth service\cite{b14}. MyAccessID has already been deployed for the EuroHPC LUMI user management project called Puhuri\cite{b15}. According to NIST, Zero Trust (ZT) is the term for an evolving set of cybersecurity paradigms that move defences from static, network-based perimeters to focus on users, assets, and resources. A zero-trust architecture (ZTA) uses zero trust principles to plan enterprise infrastructure and workflows\cite{b16}.  Open source and open standard technologies have been adopted throughout the architecture, design and implementation for Isambard DRIs so that it can be extended to cover existing and future international research resources federation, while complying with institutional, domain and national regulatory and compliance frameworks.

We present the architecture, design and implementation details of the federated IAM solution, which is coupled with multiple layers of security controls, multi-factor authentication (MFA), cloud-native protocols and time-limited role-based access controls (RBAC), that has been co-designed and deployed for the Isambard-AI and HPC DRIs.  Strong identity vetting via implicit and explicit trust levels is a core design principle.  Users can use MyAccessID as a federated, trusted Identity Provider (IdP) proxy.  Then, there is an ``Identity Provider of Last Resort'' (for anyone not in MyAccessID), and a separate IdP for administrator identities.  In fact, identity vetting is interleaved with authorisation workflows such that identity registration is led by authorisation. Three levels of RBAC have been introduced at the identity management layer depending on the level of access: Researcher, Principle Investigator (PI), and Administrator. The resulting solution is deployed and in operation on the Isambard-AI phase 1 system at the time of writing\cite{b17}.

The outline of the paper is as follows: section~\ref{sec:background} details background and motivation for the work. Section~\ref{sec:design} presents design and implementation details, which is followed by usage analysis and experience via user stories in section~\ref{sec:results}. This section contains a discussion of strengths and shortcomings. Finally, we present opportunities for future work in section~\ref{sec:conclusions}.

\section{Background and Motivation}
\label{sec:background}
\subsection{UK Future of Computing Report and DRI Investment}

In Spring of 2023, an influential report was published in the UK titled ``The Future of Compute''\cite{b19}. Several years in the making, this report included a number of important and influential recommendations, including one to establish a new AI Research Resource, or AIRR, for the UK. The report recognised the lack of AI-capable supercomputing facilities in the UK, and the impact this was having on AI research, as well as AI-enabled scientific discovery. In response, the UK Government announced a £900M fund to radically boost the UK’s supercomputer capabilities.  Isambard-AI was funded to host a new UK AIRR as national DRI.

UKRI defines the building blocks of the DRI systems as:

\begin{itemize}
    \item large-scale compute facilities, including high-throughput, high-performance, and cloud computing
    \item data storage facilities, repositories, stewardship and security
    \item software and shared code libraries
    \item mechanisms for access, such as networks and user authentication systems
    \item people: the users, and the experts who develop and maintain these powerful resources
\end{itemize}

Access controls, authentication, and cybersecurity were explicitly mentioned in addition to the resources.  This is because the stated vision is a coherent state-of-the-art national digital research infrastructure that will seamlessly connect researchers, policymakers and innovators to the computers, data, tools, techniques and skills that underpin the most ambitious and creative research. Isambard DRIs aim to achieve this objective for the UK's AI research communities by undertaking the work we present in this paper.

\subsection{Federated IAM for EuroHPC and European Open Science Cloud (EOSC)—MyAccessID}
The MyAccessID IAM service is provided by GÉANT. GÉANT is the collaboration of European National Research and Education Networks (NRENs) that delivers an information ecosystem of infrastructure and services to advance research, education, and innovation on a global scale in partnership with national services such as Internet2 in the USA. One such famous global service is eduroam.  MyAccessID aims to offer a common Identity Layer for Infrastructure Service Domains (ISDs). The UK Access Management Federation (UKAMF), which is part of JISC (the UK's NREN) has deployed an evaluation platform as part of this effort\cite{b20}. It already includes a number of IdPs across the world.  The minimum requirement is Research and Scholarship (R\&S) compliance\cite{b21}. MyAccessID is based on the AARC phase 2 (AARC2) Blueprint Architecture (BPA), which is a set of software building blocks that can be used to implement federated access management solutions for international research collaborations.
 
MyAccessID has been used for providing authentication services to LUMI EuroHPC project and Fenix research infrastructure for the Human Brain Project\cite{b22}. The Authentication and Authorisation Infrastructure (AAI) proxy service of MyAccessID connects Identity Providers (IdPs) from eduGAIN, specific IdPs which are delivered in context of ISDs such as HPC IdPs, eIDAS eIDs and potentially other IdPs as requested by ISDs. The eduGAIN inter-federation service connects identity federations around the world, simplifying access to content, services and resources for the global research and education community. eduGAIN comprises over 80 participant federations connecting more than 8,000 Identity (IdPs) and Service Providers (SPs). One of the main challenges for eduGAIN is the lack of features for controlling assurance and trust from IdPs (strong identity vetting needs from some SPs, such as HPC centres).

MyAccessID delivers the Discovery Service used during the user authentication process for users to choose their IdP. It enables the user to register an account in the Account Registry, to link different identities, and it guarantees the uniqueness and persistence of the user identifier towards connected ISDs. ISDs include HPC and DRI service providers. Further development of MyAccessID has been aligned with the European Open Science Cloud (EOSC) AAI task force.  Both are being evolved as part of European and international committees in an agile manner.

\subsection{Zero Trust Architecture (ZTA)}
In response to large-scale attacks a few years ago, US executive order 14028 outlines measures for government agencies to enhance cybersecurity through a variety of initiatives related to the security and integrity of the software supply chain\cite{b23}. Other international governments have followed suit, including the UK.  Among key measures, ZTA is an important requirement. The NIST ZTA publication mentions seven tenets stating, ``A zero trust architecture is designed and deployed with adherence to the following zero trust basic tenets'' :

\begin{enumerate}
    \item All data sources and computing services are considered resources.
    \item All communication is secured regardless of network location.
    \item Access to individual enterprise resources is granted on a per-session basis.
    \item Access to resources is determined by dynamic policy—including the observable state of client identity, application/service, and the requesting asset—and may include other behavioural and environmental attributes.
    \item The enterprise monitors and measures the integrity and security posture of all owned and associated assets.
    \item All resource authentication and authorization are dynamic and strictly enforced before access is allowed.
    \item The enterprise collects as much information as possible about the current state of assets, network infrastructure and communications and uses it to improve its security posture.
\end{enumerate}

Typically, supercomputing environments are not architected for ZTA and instead focus on a trusted access and network domain within which close-to-metal performance and scalability is guaranteed.  These ZTA principles are however the drivers of Isambard DRI decisions, including the definition of research resources, least privilege access model (no blanket authorisation), managing trust, time limited sessions, monitoring, and a comprehensive identity, access and credential management system for accessing different resources (AI and HPC).  As a national research resource, a balanced approach is taken to enforce re-authentication and re-authorization as per the policy (time-based, new resource requested, resource modification), balancing security, availability, usability, and cost-efficiency.

\section{Design and implementation}
\label{sec:design}
\begin{figure*}[htbp]
\centerline{\includegraphics[width=\textwidth]{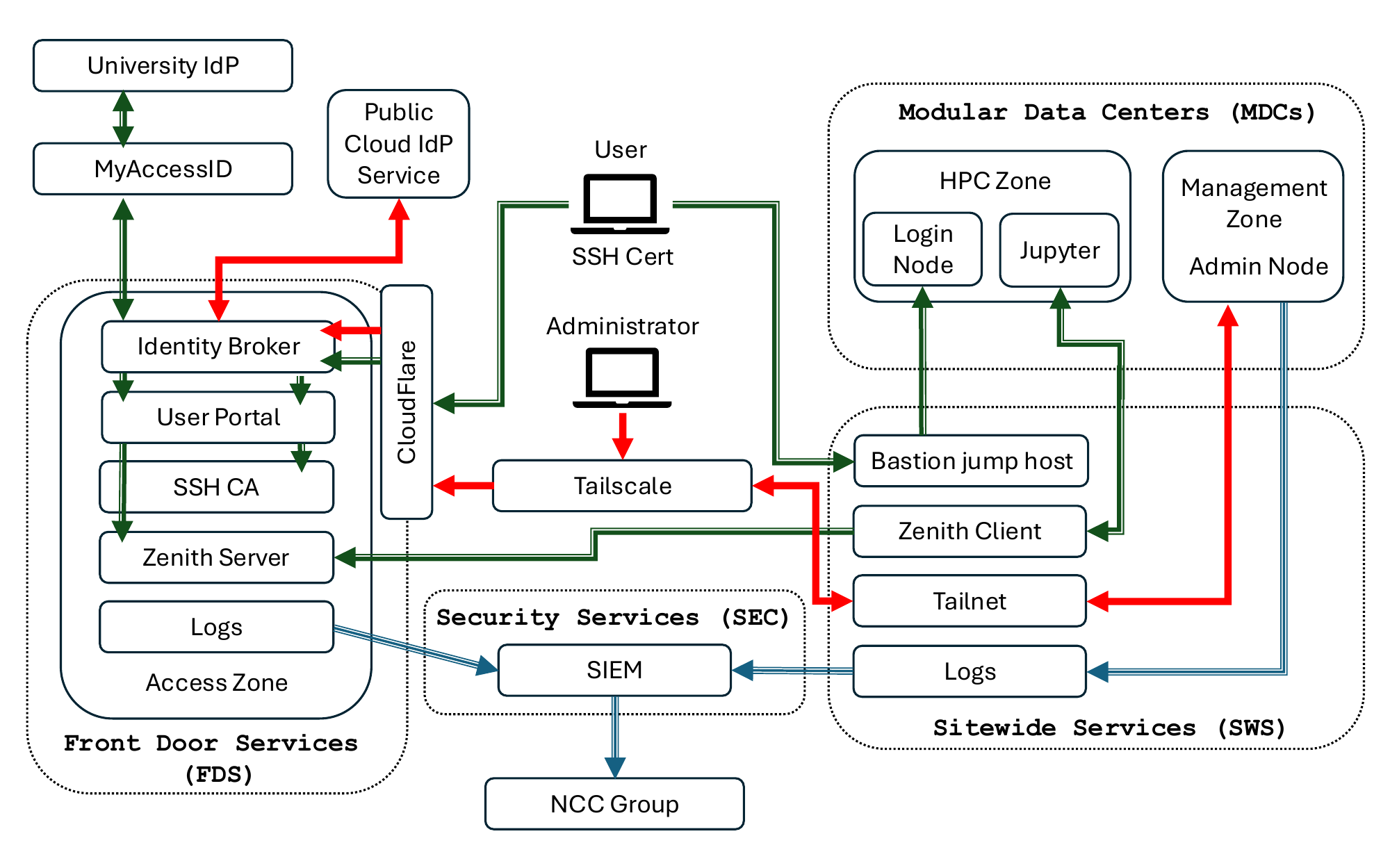}}
\caption{Implementation details of federated SSO and ZTA for Isambard DRIs, demonstrating different domains (MDCs, SWS, FDS and SEC), different zones (Access, HPC and Management) and segmentation, technical and tooling details, and workflows for user roles (project owners or PIs and researchers) and different administrator roles. RBAC is not global and is managed per service; access is controlled on a per session basis.}
\label{fig1}
\end{figure*}

The overall design and implementation is shown in \figurename~\ref{fig1}. The design uses the following principles:

\begin{itemize}
    \item All authentication and access is based on short-lived role-based access tokens (RBACs).  
    \item The Access Zone is physically and logically separated from the Management, Data Storage and High Performance Computing Zones. Only the Access Zone is directly internet accessible, and all connections to the other Zones must be authenticated using RBAC tokens generated from within the Access Zone.
    \item The Management Zone is logically separated from all other Zones. Access to the Management Zone can only be via authenticated Administrator identities adopting time-limited administrator roles. The Management Zone is not directly accessible from any other Zone or from the public internet.
    \item Security monitoring and alerting is in a new Security Zone that is physically and logically separated from all other Zones. Access is only via authenticated Administrator identities adopting time-limited security roles.
    \item Widely used, open protocols are used where available.
    \item Open source, but commercially supportable software is used where available.  
\end{itemize}

Isambard DRIs have four operating domains with fully segmented networks to achieve this design. These domains are:

\begin{description}
\item[Modular Data Centres (MDCs)] are containerised modular data centres in which Isambard-AI Phase 1, Isambard 3 and (soon to be) Isambard-AI Phase 2 are housed. They contain multiple isolated, dedicated networks for running the High Performance Computing, Management and Data Storage Zones.

\item[Sitewide Services (SWS)] are hosted in a data centre in the National Composites Centre, next to the Isambard Modular Data Centres. Sitewide Services manage the physical fibre network connections to the Modular Data Centres, and the fibre network connections to the wider internet.

\item[Front Door Services (FDS)] are hosted in public cloud and run all services related to the Access Zone. This includes connecting to external IdPs for user authentication, generation of RBAC tokens, and provision of internet-facing endpoints through which web applications running in the High Performance Computing Zone can be reached via reverse tunnels routed via clients in SWS.

\item[Security Services (SEC)] is hosted in public cloud in a different cloud account to FDS. This runs the Security Zone, hosting the Security Information Event Management (SIEM) service, as well as the interface with an external commercial 24/7 threat monitoring service provided by NCC Group.
\end{description}

In addition to the above, there are external services and dependencies. We describe the design and function of each element shown in \figurename~\ref{fig1} in these domains for achieving a federated SSO and zero trust compliant implementation. The privileged access to domains, and services within a domain, are controlled by RBAC i.e. there is no such concept as a global admin or root on all services or in all Zones. All data flowing through the design shown in \figurename~\ref{fig1} is encrypted.

\subsection{Isambard Modular Data Centres (MDCs)}
Isambard-AI phase 1 (168 Grace-Hopper superchips) and Isambard 3 (368 Grace-Grace superchips) are housed in a next-generation HPE Performance Optimised Data centre (POD) as a complete ``Self-Contained Unit (SCU)'' including all necessary power and cooling infrastructure. An MDC comprises an IT module (DC20) that includes a row of IT racks and a hot corridor (supercomputer equipment including servers, networking, and storage).  The main components of the architecture are user login nodes (user plane) and admin nodes (management plane). The services are managed with RBAC tokens and follow the model of time-limited sessions where new tokens are issued for reauthorisation.  User login nodes are essentially the gateway to the supercomputing system running essential services such as Slurm (job management and resource scheduler) and Jupyter notebook spawner services. These are not directly internet accessible. Access to these is via either a locked-down transparent SSH bastion running in SWS, or via secure Zenith\cite{b32} reverse tunnels to web accessible end-points running in FDS (described later). Access depends on multi-factor authentication using time-limited tokens. For example, access to a User login node requires a time-limited, signed SSH certificate that is generated in FDS. This certificate grants login access for a limited time period to the User login node, using the bastion in SWS as a transparent jump host.

Administrator access is completely isolated, via a separate management network, and separate administrator accounts and roles. Access to the management network is routed via SWS using Tailscale tailnets\cite{b25}. These are virtual private networks based on the open source WireGuard protocol\cite{b26}. Access to the tailnet is gated via RBAC tokens generated in FDS via a separate administrator account identity provider (described later).

All authentication takes place in FDS, meaning that there is no unauthenticated network access to anything running in SWS or the MDCs. This, coupled with extensive use of RBAC tokens and separated networks and roles, means that, essentially, a non-authorised user of a service cannot access the AI and HPC resources.

\subsection{Isambard Sitewide Services (SWS)}
MDCs for Isambard DRIs are located in a purpose-built compound at the National Composite Centre (NCC) on the edge of the City of Bristol. This was already identified as a location with sufficient power, cooling, networking and heat reuse, and being on a science park, it also met the criteria of having rapid turnaround for planning permission applications. The NCC is a digital first facility with its own data centre, that, while non HPC, has a fibre network connection to Jisc (the UK's NREN). Additional, air-cooled cabinets were installed within the NCC's data centre for running Sitewide Services and network connectivity to the MDC.

SWS is managed as a Virtual Machine (VM) service, operating primarily three sets of services.

The first is the bastion service. This comprises a redundant set of bastion jump hosts, configured as a high-availability set of VMs that are fully locked down. Their only purpose is to handle SSH connections to the jump host. They provide a transparent SSH jump host that routes SSH connections from the internet to the appropriate login nodes of the the AI and HPC login services running in the MDCs. These bastions are the only internet accessible service running in SWS, with port 22 being the only opening in the firewall between SWS and the public internet. The bastions use read-only images, and contain only the software needed to function as SSH jump hosts. SSH configs are provided to users that make use of the jump host transparent via an SSH \texttt{ProxyJump} rule\cite{b27}.

The bastions are operated as a high-availability VM set so that they can be patched and updated quickly against known security vulnerabilities. Load balancing within the VM set provides fault tolerance, and allows live updates to be undertaken without risk of disruption to the service. The design also makes implementation of an externally managed ``kill switch'' easier in case of a threat and attack, without waiting for a direct intervention from the Isambard team. In such a case, SSH access to flagged users can be terminated and blocked, thereby immediately severing that user’s connection. Or, in extreme cases, the entire bastion service could be shut down, blocking access to all.

The second function of SWS is to provide the log gathering services and monitoring from all resources/assets inside the Isambard MDCs. This includes all system specific logs from the HPE environments that gather server, network and storage usage data, and environmental monitors such as the Data Centre Inventory Manager (DCIM). These are managed to support easy administrator monitoring of the service, as well as collection and forwarding of logs to SEC for security monitoring and threat detection. Data sent to SEC is combined with data collected by log forwarders running on the bastions and on the login nodes to be ingested by NCC Group’s 24/7 external threat monitoring and alerting service. The log forwarders are managed as an external services outside the Isambard team by the University of Bristol central IT services.  They ingest a limited amount of data that has been agreed with the University’s security team.

The third function of SWS is to provide access to the administrator services and networks needed to operate the Isambard DRIs. This is via Tailscale tailnets\cite{b25}, with tailscale end-points operating across VMs in SWS and management nodes in the MDCs. Access to the tailnets is via RBAC tokens obtained from FDS. Access to management services within the tailnets follows an RBAC model with additional tokens and roles. Similar to the bastions, there is an externally managed kill switch for the management tailnets, for rapid intervention in case of emergencies where decisions need to be taken and enforced immediately to contain a threat, or stop a threat from spreading.

\subsection{Front Door Services (FDS)}
Front Door Services provides the Access zone. With the exception of the jump host, it is the only internet-accessible part of the implementation. To minimise the blast radius, FDS is physically and logically separated from the MDCs and SWS. It is hosted in public cloud. All services are hosted in a private Kubernetes cluster placed within a Virtual Private Cloud (VPC). This VPC is not directly internet accessible. Instead, services running on the Kubernetes cluster are exposed via Cloudflare zero-trust reverse tunnels\cite{b28}. Cloudflare tunnels provide an extremely high level of protection, mitigating distributed denial of service (DDoS) attacks and automatically blocking access that Cloudflare has determined to be a threat.

The central service running in FDS is an identity broker. It authenticates users via external Identity Providers (IdPs), and then generates RBAC tokens using those authenticated identities. This identity broker is used to connect to MyAccessID to authenticate users, and to generate the RBAC tokens that grant access to services in SWS and the MDCs.

Administrator identities are authenticated via the public cloud's IdP. Public cloud provides IdP as a managed service. These provide strong guarantees on use of multifactor authentication (MFA), ensuring that administrator identities are authenticated via hardware key MFA tokens. The identity broker uses these authenticated administrator identities to generate RBAC tokens that provide administrator access.

In addition, a separate public cloud IdP instance is provisioned as an ``Identity Provider of Last Resort''. This is used to authenticate users who cannot use MyAccessID, e.g. because they work for a company or other institution.

FDS also hosts the Isambard user and project management portal. This provides a user-friendly web interface for managing projects and user access, based on a user’s role. For example, a user in the Principle Investigator (PI) role can invite other users to join a project in Researcher roles. This would grant those Researchers access to the Isambard DRI services that are connected to the PI’s project. The user portal provides an API to query the roles and level of access of a user. This is used as part of the identity broker's login flows to generate the appropriate time-limited RBAC tokens to grant access to users to the services they are requesting.

For example, FDS hosts a SSH certificate authority (CA) which is used to generate time-limited SSH certificates to let users connect to login nodes in the MDCs via the jump host running in SWS. A user initiates a signing request by running a SSH certificate client application on, e.g. their laptop. This passes the user’s SSH public key to the client application, which initiates a login flow whereby the user is directed to the identity broker web page. Here, the identity broker authenticates the user, the portal asserts that access is permitted, and the identity broker is provided with the list of project-specific Linux user accounts associated for that user for all of their active projects. This information is routed from the identity broker to the SSH CA, which signs the user’s public key, generating a time-limited certificate with this information and necessary attributes encoded. The client program receives the certificate, and then (optionally) updates the user’s SSH configuration file to include aliases for all of their projects and a \texttt{ProxyJump}\cite{b27} command to route access to the login nodes via the jump host. The user can then SSH connect to the login node using that alias, e.g. \texttt{ssh project-name.example.cluster} would connect to the login node of the cluster using the user’s Linux account associated with the project called ``project-name''. Details of the user’s Linux account and use of the jump host is transparent.

For web services, FDS hosts a Zenith server\cite{b32}. This provides authenticated internet end-points for web services running in the MDCs. For example, \texttt{https://example.com/jupyter} could be the end-point for Jupyter running on Isambard-AI Phase 1. This URL can be accessed only by authenticated identities. A user navigating to this URL triggers an identity broker login flow that authenticates their identity, and connects to the user portal to verify access to the web service. If successful, this generates a time-limited RBAC token that is passed as a HTTP header to the Jupyter authenticator running in the MDCs to grant access. The Zenith server routes traffic to a Zenith client connected to the Jupyter service via a reverse tunnel from the MDCs to FDS via SWS. For additional security, the Zenith server end-point is placed behind a Cloudflare zero-trust reverse tunnel, thereby providing DDoS protection. Connections via Zenith are logged to SEC, with a kill-switch implemented as for other services.

\subsection{Security Services (SEC)}

Isambard has a virtual central Security Operations Centre (SOC), running. It is hosted in public cloud, following the design laid out in the AWS Security Reference Architecture\cite{aws_sra}.

The SOC performs three main tasks: 1) aggregate and scan logs from across MDCs, SWS and FDS to identify potential attacks and raise alerts, 2) inventory all virtual machines in SWS and FDS to track software versions for vulnerabilities, and 3) provide security configuration assessment to aid with compliance with best-practice guidelines, such as CIS\cite{cis}.

\section{Results and analysis}
\label{sec:results}
In this section, we walk through the implementation through different user stories and the resulting workflows.  We analyse the strengths and shortcomings for federated single sign-on and zero-trust security architecture. We demonstrate the steps taken for identity verification, establishing device trust principles, segmentation of different networks to reduce the attack surface, and enforcement of policies through access controls.

\begin{figure}[htp]
\centerline{\includegraphics[width=0.42\textwidth]{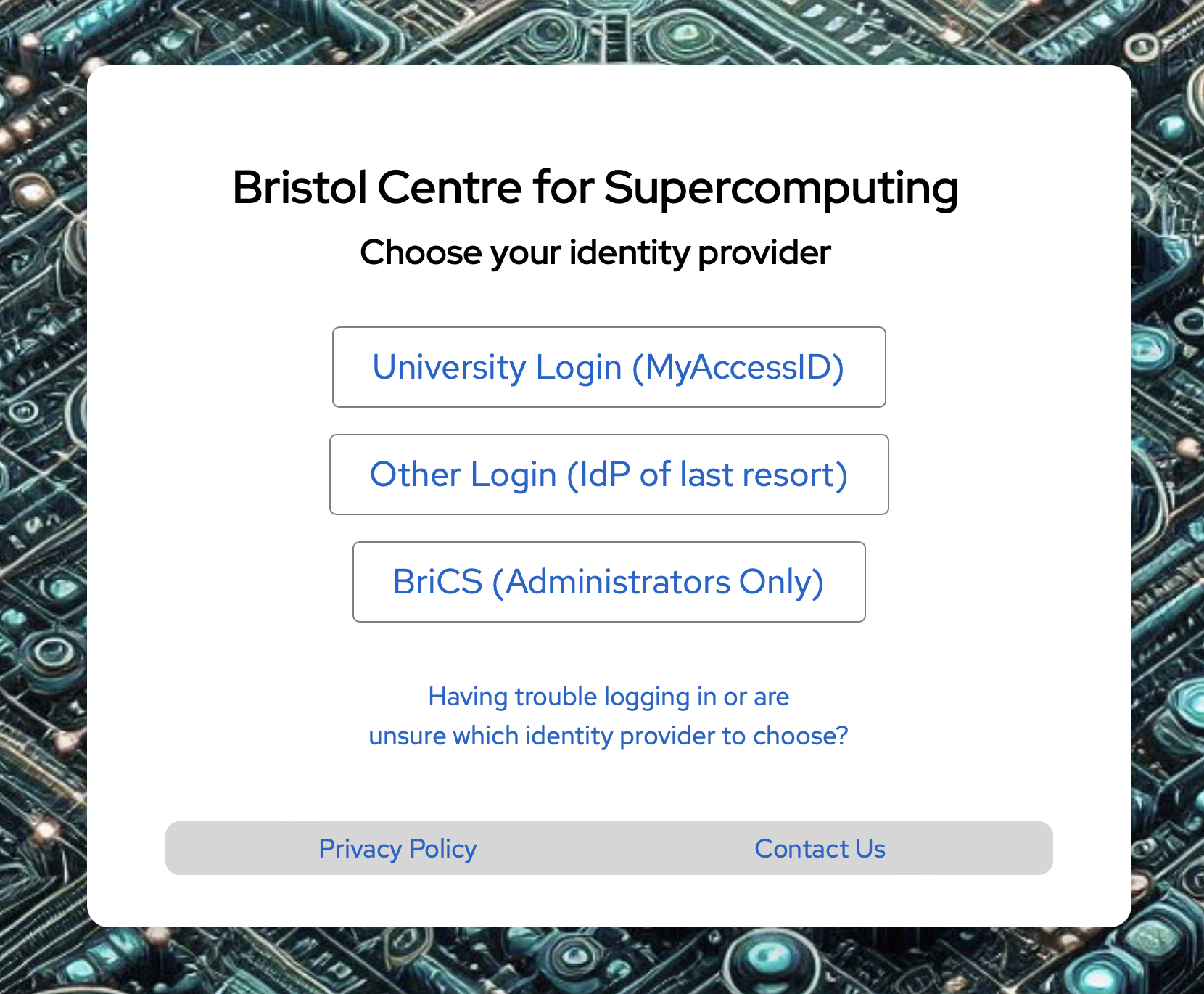}}
\caption{Login page for the Isambard services. Users choose their identity provider, which for most researchers would be ``University Login (MyAccessID)''. This also links to the privacy policy, route to gain help with logins and methods of contacting the technical team.}
\label{fig2}
\end{figure}

The login screen for Isambard services (\figurename~\ref{fig2}) is the first gateway for users where they choose their identity provider and accept the usage terms and conditions and acceptable usage policies. This page also lists information security, data protection and privacy policies that govern Isambard DRI resources. The user of the services is expected to login with their federated, academic and research identities (MyAccessID).  An identity of last resort is available to users whose institutional IdPs are not part of MyAccessID. For instance, vendors or government entities, including the UK AI Safety Institute, are not part of MyAccessID federation.  This solution can be extended to other trusted IdP federations. The last option is limited to the Isambard team members who would need privileged access to services such as the Isambard portal and service desk.

We followed an agile, DevOps approach for implementation where we defined a few user stories.  A subset of these user stories are listed here.  A user has a different meaning in an HPC system context i.e. someone who accesses the cluster, compilers applications and submits jobs.  Here we distinguish these by roles in order to implement role-based access controls (RBAC) and, as needed, additional controls for privileged accesses.

\subsection{User stories}
\subsubsection{A project owner or principle investigator (PI) who has been granted an allocation is invited to join the project}
Two roles are involved here.  One is an allocator role, which accesses the user and project management portal (running on public cloud) via admin authentication and authorisation.  The allocator creates a project, sets resources for authorising the user as a Principal Investigator (PI) role—the second role.  The PI accesses the portal through federated ID (MyAccessID). If the PI's institution is a trusted member of the federation, the PI logs in and accepts Terms and Conditions for usage and Privacy Policy. The authentication is led by authorisation.  For instance, if a person has not been granted a role of a PI and is not in the access control list with authorisation to access resources, the registration process will fail after the MyAccessID registration. The PI can use an identity of last resort (managed by the Isambard team using public cloud managed IdP) if the PI's institution is not part of MyAccessID federation. Each project is time and resource limited.  Access is revoked after expiration or on-demand. All information related to the project including the PI and users is removed from the authorisation list.

\subsubsection{A BriCS admin registers an administrators only account}
We designed the solution with an assumption that this group is small, around 20 individuals, and are legally part of the same institution i.e. the Bristol Centre for Supercomputing (BriCS) at the University of Bristol, primarily for the coverage of data protection and privacy policies.  With this setup, a new admin role is invited to create an account via the public cloud managed IdP service for setting up a hardware-token MFA controlled account. There is at least one human check involved in confirming the identity and updating AWS identity centre (AWS admin group is a small subset of BriCS' admin roles). The admin access does not provide global access to all Isambard services.  There are authorisation and access controls at individual service levels. Access is revoked when an individual leaves the group.

\subsubsection{A cluster user sets up an account}
This workflow is typically triggered when a ``Researcher'' role invitation is triggered by a PI. A user receives an email to register for an account and project. The remaining steps are similar to user story 1 except a researcher has fewer functions and visibility compared to a PI. A researcher cannot invite other researchers. A PI can revoke or remove a researcher from a project, which removes their user authorisation. Authorisation is linked to a trusted identity.  Authentication will fail if a user is no longer affiliated with the organisational IdP.

\subsubsection{A cluster user connects via SSH to the AI platform}
This involves multiple steps. As part of the user stories 1 and 3, a researcher selects a Linux user name. The next step is downloading and running the SSH certificate client application to a local device to authorise the user’s access via their chosen public key. The client application initiates the login flow as described previously and signs the certificate. The returned SSH certificate has a short valid session time, after which a new certificate must be generated. A unique UNIX username is generated for each user’s access to each project to ensure ZTA resource access requirements. This detail, and connection via the jump host, is hidden from the user by aliases that can (optionally) be added to the user’s SSH configuration file by the SSH certificate client application.

\subsubsection{A system administrator performs a privileged operation}
Here we highlight multiple layers of authentication and authorisation in practice for an admin role who is authorised to access the management plane of the cluster to perform privileged operations. There is a separate access control list on the cluster level and additional controls.  The user registration is similar to user story 2 except there is a separate bastion dedicated for accessing the cluster management plane, and connection to this is only via logging into the Tailscale administrators’ tailnet. Essentially, as a zero-trust architecture requirement, it establishes segmentation and enforces policies at each level for accessing the management plane of a cluster.

\subsubsection{A cluster user connects to a Jupyter notebook}
Web applications can be published via secure URLs. A user navigates to the Jupyter URL, e.g. \texttt{https://example.com/jupyter}. This takes them through a identity broker login flow that asserts their identity and checks with the user portal that the user has access to the application. If successful, network traffic from the user to the Jupyter URL is routed via a reverse tunnel to the Jupyter authenticator running on the cluster login node, passing the authentication token with access attributes as an HTTP header. The Jupyter authenticator validates this token against the OpenID Connect endpoint from the identity broker in FDS. If successful, a Jupyter user session is spawned on a compute node and the researcher able to work with their data and code via a browser-based notebook.  This process is seamless to the user within an authorised web session.

\subsection{Agile development analysis}
An agile approach using the user stories was highly effective in identifying strengths and shortcomings of design choices and tools that are used for implementation. It enabled the rapid development and deployment of the architecture in time for both production use of BriCS services, and the running of a successful workshop at the Research Software Engineering Conference 2024 (RSECon24) in September\cite{b33}. The conference tested the Jupyter notebook user story at scale, with 45 trainees logging in and running notebooks simultaneously. The main advantage and feedback was the positive experience of the cloud look-and-feel of the process and workflows, which was appreciated (for the most part except for individuals who are only familiar with classic HPC cluster access). There are several shortcomings though that we intend to address in the future. The first issue is that the zoning and segmentation can be confusing, and will have scaling challenges as we introduce additional services and users.  We need to grow a DevSecOps culture in the team and beyond, to establish and harden these practices. External tools such as Tailscale are highly effective and efficient. However, these have not been widely deployed in HPC environments. This may require additional development efforts for integrating with HPC system administration workflows. Another shortcoming of our solution is reliance on a widely used but limited AAI federation for R\&D institutions. The public cloud managed IdP of last resort service does not provide federation. There should be international efforts to address this bottleneck because this requires a wider agreement on established IdP level of trust/level of assurance. Finally, the usage of encryption, while widespread for the implementation of authentication and authorisation workflows and tools, is not yet as widely adopted in the HPC ecosystem, such as in the high performance network and parallel file systems.  There are performance and functional concerns that are beyond the scope of this paper.

\section{Conclusions and future work}
\label{sec:conclusions}
We have presented an approach for federated single sign-on and zero trust architecture for the Isambard Digital Research Infrastructure (DRI) that are part of an AI research resource federation for UK research and academia, as well as government entities.  This approach is governed by the ever evolving cybersecurity compliance and regulatory landscape, where we balance lowering access barriers for users while ensuring access controls. An ability to mitigate risks and threats is a key design requirement. The work presented in this paper was completed in less than 6 months. The Isambard-AI phase 1 system arrived at the end of March 2024. We started with a small dedicated team and with limited legacy Isambard 2 operations, which proved to be highly advantageous.

We demonstrated adoption of an open standards and open source platform for identity control and vetting. In addition, we adopted the key tenets of the zero-trust architecture. It starts with identifying zones and mapping dependencies of key services, data and applications within each zone. We identified three classes of accesses for segmentation and implemented MFA based authentication methods to reflect sensitivity and business criticality. Segmentation of network domains allowed us to isolate and contain different threats. End-to-end connectivity of users and devices is visible for threat monitoring and access controls. Encryption is applied for all IAM workflows. We are in the process of gradually introducing a DevSecOps culture in the team to adapt to technical and regulatory challenges. Through different user stories, we demonstrated implementation and workflows.

Our next steps is to achieve CAF compliance for the baseline profile. This will be done as an independent assessment and will require hardening of tools as well as processes and procedures. We then plan on ISO 27001. Encryption and multi-tenancy on the HPC system level are in progress, as well as increased telemetry needed for introducing DevSecOps.

\section*{Acknowledgments}
Isambard-AI is funded by the UK Government’s Department of Science, Innovation and Technology (DSIT) via UK Research and Innovation (UKRI) and Science and Technology Facilities Council (STFC) [grant number ST/AIRR/I-A-I/1023]. This work was additionally funded via UKRI projects [grant numbers EP/X034828/1, EP/Z002672/1]. The work presented in this paper has only been possible thanks to a very talented and extremely hardworking team of people across the University of Bristol, JISC IAM, GÉANT MyAccessID and the Waldur OpenNode teams.

\printbibliography

@online{b1,
	title = {{Digital research infrastructure}},
	author = {{UKRI}},
	url = "https://ukri.org/what-we-do/creating-world-class-research-and-innovation-infrastructure/digital-research-infrastructure",
	year = 2024,
}

@online{b2,
	title = {{Destination Earth}},
	url = "https://digital-strategy.ec.europa.eu/en/policies/destination-earth",
	year = 2024,
}

@online{b3,
	title = {{DARE UK}},
	url = "https://dareuk.org.uk/",
	year = 2024,
}

@online{b4,
	title = {{Introducing the AI Safety Institute}},
	url = "https://gov.uk/government/publications/ai-safety-institute-overview/introducing-the-ai-safety-institute",
	year = 2023,
}

@inproceedings{b5,
	title = {{Isambard-AI: a leadership class supercomputer optimised specifically for Artificial Intelligence}},
	author = "Simon McIntosh-Smith and Sadaf R. Alam and Christopher Woods",
	booktitle = "Proceedings of Cray User Group (CUG) conference",
	year = 2024,
    % waiting for publication to get a DOI
}

@article{b6,
	author = {Sadaf R Alam and Miguel Gila and Mark Klein and Maxime Martinasso and Thomas C Schulthess},
	title ={Versatile software-defined HPC and cloud clusters on Alps supercomputer for diverse workflows},
	journal = {The International Journal of High Performance Computing Applications},
	volume = {37},
	number = {3-4},
	pages = {288-305},
	year = 2023,
	doi = {10.1177/10943420231167811},
}

@online{b7,
	title = {{NVIDIA NGC}},
	url = "https://nvidia.com/en-gb/gpu-cloud/",
	year = 2024,
}

@online{b8,
	title = {{The rise of Dawn}},
	url = "https://cam.ac.uk/stories/ai-supercomputer-dawn-research-energy-medicine-climate",
	year = 2024,
	month = feb,
}

@online{b9,
	title = {{Cyber Assessment Framework}},
	author = {{NCSC}},
	url = "https://ncsc.gov.uk/collection/cyber-assessment-framework",
	year = 2024,
}

@online{b10,
	title = {{Government Security Classifications Policy}},
	url = "https://gov.uk/government/publications/government-security-classifications/government-security-classifications-policy-html",
	year = 2024,
}

@misc{b11,
	title = {{High-Performance Computing Security: Architecture, Threat Analysis, and Security Posture}},
	author = "Yang Guo and others",
	doi = "10.6028/NIST.SP.800-223",
	year = 2024,
}

@online{b12,
	title = {{AARC}},
	url = "https://aarc-community.org",
	year = 2024,
}

@online{b13,
	title = {{MyAccessID}},
	url = "https://wiki.geant.org/display/MyAccessID",
	year = 2024,
}

@inproceedings{b14,
  author = {Tuecke, Steven and Ananthakrishnan, Rachana and Chard, Kyle and Lidman, Mattias and McCollam, Brendan and Rosen, Stephen and Foster, Ian},
  booktitle = {2016 IEEE 12th International Conference on e-Science (e-Science)}, 
  title = {Globus auth: A research identity and access management platform}, 
  year = 2016,
  pages = {203-212},
  doi = {10.1109/eScience.2016.7870901},
}

@online{b15,
	title = {{Puhuri Architecture}},
	url = "https://puhuri.io/architecture",
	year = 2024,
}

@online{b16,
	title = {{Zero Trust Architecture}},
	author = {{NIST}},
	url = "https://nist.gov/publications/zero-trust-architecture",
	year = 2020,
}

@online{b17,
	title = {{Getting started on Isambard-AI}},
	author = {{BriCS}},
	url = "https://docs.isambard.ac.uk/user-documentation/getting_started/",
	year = 2024,
}

@online{b19,
	title = {{Independent Review of The Future of Compute: Final report and recommendations}},
	author = "Zoubin Ghahramani and others",
	url = "https://gov.uk/government/publications/future-of-compute-review/the-future-of-compute-report-of-the-review-of-independent-panel-of-experts",
	year = 2024,
	month = mar,
}

@online{b20,
	title = {{UK Access Management Federation for Education and Research}},
	url = "https://ukfederation.org.uk",
	year = 2024,
}

@misc{b21,
  author       = {Harris, Nicole},
  title        = {{REFEDS Research and Scholarship Entity Category}},
  month        = jul,
  year         = 2022,
  publisher    = {Zenodo},
  version      = {1.3},
  doi          = {10.5281/zenodo.6832218},
}

@article{b22,
	author = {Alam, Sadaf R. and Bartolome, Javier and Carpene, Michele and Happonen, Kalle and Lafoucriere, Jacques-Charles and Pleiter, Dirk},
	title = {Fenix: a Pan-European federation of supercomputing and cloud e-infrastructure services},
	year = {2022},
	issue_date = {April 2022},
	publisher = {Association for Computing Machinery},
	address = {New York, NY, USA},
	volume = {65},
	number = {4},
	issn = {0001-0782},
	doi = {10.1145/3511802},
	journal = {Commun. ACM},
	month = mar,
	pages = {46–47},
	numpages = {2},
}

@online{b23,
	title = {{Executive Order 14028: Improving the Nation's Cybersecurity}},
	url = "https://gsa.gov/technology/it-contract-vehicles-and-purchasing-programs/information-technology-category/it-security/executive-order-14028",
	year = 2021,
}

@online{b25,
	title = "What is a tailnet?",
	url = "https://tailscale.com/kb/1136/tailnet",
	year = 2024,
}

@inproceedings{b26,
	title = {{WireGuard: Next Generation Kernel Network Tunnel}},
	author = "Jason A. Donenfeld",
	booktitle = {{NDSS}},
	url = "https://wireguard.com/papers/wireguard.pdf",
	year = 2017,
	month = feb,
}

@online{b27,
	title = {{SSH to remote hosts through a proxy or bastion with ProxyJump}},
	url = "https://redhat.com/sysadmin/ssh-proxy-bastion-proxyjump",
	year = 2019,
	month = dec,
}

@online{b28,
	title = {{Argo Tunnel: A Private Link to the Public Internet}},
	url = "https://blog.cloudflare.com/argo-tunnel",
	year = 2018,
	month = 4,
}

@software{b32,
	title = "Zenith",
	url = "https://github.com/azimuth-cloud/zenith/",
}

@unpublished{b33,
    title = {{Introducing Isambard-AI: new leadership-class UK facility to advance AI}}, 
    author = "Matt Williams and Christopher Woods and Paul Graham and Karin Sevegani",
    year = "2024",
    note = "RSECon24"
}

@online{aws_sra,
        title = {{AWS Security Reference Architecture (AWS SRA)}},
        url = "https://docs.aws.amazon.com/prescriptive-guidance/latest/security-reference-architecture/welcome",
        year = 2024,
}

@online{cis,
        title = {{CIS Benchmarks}},
        url = "https://www.cisecurity.org/cis-benchmarks",
        year = 2024,
}

\end{document}